# Electron Doping of a Double Perovskite Flat-band System

Lun Jin[1,] *, Nicodemos Varnava[2,4], Danrui Ni[1], Xin Gui[1], Xianghan Xu[1],

Yuanfeng Xu[3], B. Andrei Bernevig[3,4,5], and Robert. J. Cava[1,] *

[1]Department of Chemistry, Princeton University, Princeton, New Jersey 08544, USA

[2] Department of Physics & Astronomy, Rutgers University, Piscataway, New Jersey 08854, USA

[3] Department of Physics, Princeton University, Princeton, New Jersey 08544, USA

[4]Donostia International Physics Center, P. Manuel de Lardizabal 4, 20018 Donostia-San Sebastian, Spain

[5]IKERBASQUE, Basque Foundation for Science, Bilbao, Spain

**\*** E-mails of corresponding authors: ljin@princeton.edu; rcava@princeton.edu

This PDF file includes:

    Main Text
    Figures 1 to 8
    Tables 1 to 2

**Abstract**

Electronic structure calculations indicate that the $Sr_2FeSbO_6$ double perovskite has a flat-band set just above the Fermi level that includes contributions from ordinary sub-bands with weak kinetic electron hopping plus a flat sub-band that can be attributed to the lattice geometry and orbital interference. To place the Fermi energy in that flat band, electron doped samples with formulas $Sr_{2-x}La_xFeSbO_6$ ($0 \leq x \leq 0.3$) were synthesized and their magnetism and ambient temperature crystal structures determined by high-resolution synchrotron X-ray powder diffraction. All materials appear to display an antiferromagnetic-like maximum in the magnetic susceptibility, but the dominant spin coupling evolves from antiferromagnetic to ferromagnetic on electron doping. Which of the three sub-bands or combinations is responsible for the behavior has not been determined.

**Significance Statement**

Flat-band electronic materials have been a keen research topic in theoretical materials physics due to their potential for hosting strongly correlated electrons and hence exotic physical phenomena. This study explores the electronic flat-band (i.e. an electronic band that has minimal variation in energy over a significant portion of the Brillouin zone) located just above the Fermi level in $Sr_2FeSbO_6$. The origin of the flat band is theoretically determined, distinguishing it from ordinary bands, and, to experimentally place the Fermi energy in that flat band, electron-doped samples are synthesized and analyzed, yielding some features that cannot be adequately described by classic physics in full. Our work shows that this doped double perovskite is an unusual electronic system worthy of further study.

# Introduction

Flat-band electronic materials have been a keen research topic in theoretical materials physics due to their potential for hosting strongly correlated electrons and hence exotic physical phenomena. However, related experiments are mostly limited to engineered materials, such as moiré system (the electron density is necessarily low due to the large unit cell), rather than naturally occurring three-dimensional stoichiometric crystalline materials which has a potential for physical properties associated with high electron density. Complex metal oxides, as a vital component of stoichiometric crystalline materials, have been of enduring interest in a wide range of disciplines, as they can exhibit physical and chemical properties that are in large part determined by their crystal structures and electron count.(1–5) The perovskite family, as one of the largest structural types in complex metal oxides, serves an important role in the material design area. B-site cation ordered double perovskites, formula $A_2BB'O_6$, although more difficult to prepare than their cation disordered analogs, can exhibit greater potential in engineering designed-for-purpose materials.(6–8)

The double perovskite $Sr_2FeSbO_6$, the basis of this study, has previously been reported and discussed regarding its crystal structure, with reports ranging from cubic (*Fm*-3*m*) to monoclinic (*P*2$_1$/*n* and *I*2/*m*) and further displaying a range of B-site cation mixing (about 80 – 90%), but without an in-depth discussion of its physical properties and band structures.(9–14) Our previous electronic structure calculations(15) found it to exhibit a nearly flat band at energies just above the Fermi level through the entire Brillouin zone, due at least in part to the lattice geometry. Those calculations are augmented here, where we find that some of the orbitals contributing to the band arise from relatively poor Fe *d*-state orbital overlap (i.e., a signature of localized Fe ion states), but that a third arises from the lattice geometry and electron interference. This doped double perovskite is thus an unusual electronic system worthy of further study due to the potential electron occupancy of its topological flat band. Flat-band materials, especially those that have topological electronic flattened bands near or at the Fermi level, are prime candidates for hosting strongly-correlated electrons(16) and can lead to exotic physical phenomena including magnetism, superconductivity, and the quantum anomalous Hall effect.(17–25) Therefore placing electrons into those bands, achieved here through the La-doping of $Sr_2FeSbO_6$, is worth investigating to determine whether unusual properties can be triggered.

In this study, the electronic structure of this material is further investigated, and to experimentally realize the flat band doping, $La^{3+}$ is chosen to partially substitute for $Sr^{2+}$ on the A-site of the Sr-Fe-Sb-O double perovskite ($Sr_{2-x}La_xFeSbO_6$), hence formally reducing some of the $Fe^{3+}$ centers to $Fe^{2+}$ centers in the B-site octahedra. The crystal structures of the undoped parent phase $Sr_2FeSbO_6$ and the newly synthesized La-doped materials are carefully examined by high-intensity high-resolution synchrotron powder X-ray diffraction. We find that a strong magnetic response is induced through the doping: the observations can be explained by both conventional and unconventional band occupancy. A delicate structure-property relationship is illustrated in the series of materials presented in this study, as evidenced by the modified properties upon introducing $La^{3+}$ dopant; changes in the crystal structure, magnetism, and electronic band occupancy of the parent compound are induced. Our experimental observations emphasize that it is critical to fully determine the crystal structures of materials prior to comprehensively understanding their physical properties. This study is a good example of how a characteristic of interest to physicists can be found in a class of materials of interest to solid-state chemists.

## Results and Discussion

**Structural Characterization**

The stacked profile of lab pXRD data (Figure 1) indicates that the maximum $La^{3+}$–for–$Sr^{2+}$ substitution level is about 30% before the sample is no longer single phase. Extremely small amounts of the impurity phases $Sr_7Sb_2O_{12}$ (PDF Card No. 00–034–1111 in ICDD) and $Fe_2O_3$ (PDF Card No. 00–013–0458 in ICDD) start to be observed in the $x = 0.3$ pattern (in too trivial an amount to be reliably quantitatively characterized by structural refinement). These non-perovskite peaks grow in the $x = 0.4$ and 0.5 patterns, while in contrast the perovskite peaks (e.g. [110] and [002]) have shrunk significantly, as highlighted in the embedded plot of Figure 1. This behavior indicates that the limit of the La solid solution is near $x = 0.3$ for our synthetic conditions.

The ambient temperature crystal structure of the undoped parent phase $Sr_2FeSbO_6$ has been reported in the literature several times, but the symmetries range from cubic ($Fm$-$3m$) to monoclinic ($P2_1/n$ and $I2/m$) with a small distribution in the degree of B-site cation ordering.(9–14) In our laboratory pXRD data, multiple weak reflections are clearly observed in addition to the strong peaks that can be indexed by a cubic unit cell (space group $Fm$-$3m$), and thus the symmetry of the unit cell is not cubic. Both monoclinic space groups mentioned above ($P2_1/n$ and $I2/m$) were refined against lab pXRD data and converged smoothly to give essentially undistinguishable agreement parameters. Although neutron and lab X-ray powder diffraction data of the undoped parent phase $Sr_2FeSbO_6$ have been reported, high-intensity high-resolution synchrotron X-ray powder diffraction, which is capable of the observation of weak reflections and small lattice parameter distortions attributed to tilts or distortions of octahedra and subtle differences in B-site cation ordering, has not been previously reported.

The subtle structural issues in this system make it appropriate to use high-resolution high signal-to-noise synchrotron X-ray powder diffraction (SXRD) to determine the crystal structure of the undoped parent phase $Sr_2FeSbO_6$ and the La-doped materials. Rietveld refinements of the host and doped materials were performed using the two monoclinic unit cells (space group $P2_1/n$ and $I2/m$). All refinements converged smoothly and provided satisfactory agreement parameters. Observed, calculated and difference plots from the Rietveld refinement of $Sr_{1.9}La_{0.1}FeSbO_6$ against the SXRD data are shown as an example in Figure 2a, and the crystal structure together with selected bond lengths for the $FeO_6$ and $SbO_6$ octahedra are depicted in Figure 2b, while the rest can be found in Figures S1-S3. In addition, it is safe to conclude that the oxygen stoichiometry is very close to the nominal value, in part due to the previous neutron diffraction studies, and, in addition, for all compositions prepared in this study based on our structural refinements against the collected SXRD data. A close inspection of the low-$2\theta$ range, shown in the two inserted plots in Figure 2a, emphasizes the differences between the two potential monoclinic space groups (space group $P2_1/n$ and $I2/m$). It is clearly observed that the Bragg reflections belonging to space group $P2_1/n$ are absent in the SXRD data collected at ambient temperature (marked by arrows in the two insets), which agrees with previously reported(14) ambient temperature neutron powder diffraction (NPD) data. The absence of primitive Bragg reflections commonly exsists in all the refinements of ambient temperature SXRD data collected for the $Sr_{2-x}La_xFeSbO_6$ ($0 \leq x \leq 0.3$) double perovskite series. Hence we assign the body-centered monoclinic unit cell (space group $I2/m$) to the ambient temperature crystal structure of the undoped parent phase $Sr_2FeSbO_6$ and the La-doped materials. As is the case for many perovskites, the monoclinic distortion of the cubic perovskite cell is primarily due to the tilting of the octahedra.

There are only two units of charge difference between the two B-site cations $Fe^{3+}$ and $Sb^{5+}$ and, although they are quite different chemically, their effective ionic radii are quite close (for high-spin $Fe^{3+}$: $r = 0.645$ Å; and for $Sb^{5+}$: $r = 0.60$ Å)(26). Thus it is widely accepted that a certain extent of anti-site disorder can exist in this material.(9–14) Therefore, we carefully examined the degree of B-site cation ordering in our Sr$_{2-x}$La$_x$FeSbO$_6$ ($0 \leq x \leq 0.3$) double perovskite series by refining the site-fraction occupancy for the two crystallographically independent B-cation sites. The refinement results show that the degree of ordering is about 95% in our undoped parent phase Sr$_2$FeSbO$_6$ ($x = 0$), which is higher than the values reported previously.(9–14) This difference can be attributed to the synthetic conditions employed in the current study: multiple high temperature (1500 °C) firings of the ceramic mixture aid the B-site cation ordering, which was also mentioned in a previous neutron diffraction study(14). In contrast, the degree of ordering in our La-doped materials ($x$ = 0.1, 0.2, 0.3) is 100%, i.e. no anti-site disorder is observed. Hence, the two crystallographically independent B-cation sites were assigned to Fe$^{3+}$ and Sb$^{5+}$ respectively, just like other fully ordered double perovskites(6, 27). The structural parameters and crystallographic positions of Sr$_{1.9}$La$_{0.1}$FeSbO$_6$ thus determined are presented in Table 1 and those of the remaining compositions can be found in Tables S1-S3.

To further confirm the success of the La$^{3+}$–for–Sr$^{2+}$ substitution in our materials, the lattice parameters for Sr$_{2-x}$La$_x$FeSbO$_6$ ($0 \leq x \leq 0.3$) were extracted from Rietveld refinements of the SXRD data and are plotted against the doping level $x$ in Figure 3. In general, the lattice parameters $a$, $b$, $c$ and cell volume $V$ all get larger, while the angle $\beta$ evolves from slightly above 90° to slightly less than 90° (90 + δ formally, not 90 - δ) with La-doping. The changes in these parameters between $x$ = 0 and 0.1 are quite sharp, while those between later compositions are much shallower, consistent with the qualitative change from partial ordering to complete ordering among the B-site cations. The effective ionic radius of La$^{3+}$ ($r$ = 1.032 Å)(26) dopant is actually smaller than the original A-site cation Sr$^{2+}$ ($r$ = 1.18 Å)(26), and hence a naive view might be that there should be shrunken lattice parameters. The unit cell expansion observed can be attributed, however, in an ionic model, to the significantly larger Fe$^{2+}$ ($r$ = 0.78 Å) centers in some of octahedra compared to the original $Fe^{3+}$ ($r = 0.645$ Å) centers. This formal introduction of $Fe^{2+}$ occurs as a consequence of the $La^{3+}$–for–$Sr^{2+}$ substitution used to raise the Femi Level of this material into the flat bands.

**Origin of the Flat Bands**

The $Sr_2FeSbO_6$ double perovskite crystallizes in space group $I2/m$; the spin structure for the calculations was set to antiferromagnetic based on our experimental data and the literature(12, 14). The calculated band structure and electronic density of states (DOS) of the undoped parent phase $Sr_2FeSbO_6$ are depicted in Figure 4a. The states close to the Fermi level arise from Fe-$d$, and O-$p$ states while the Sb-s states lie higher in energy. Due to the octahedral environment, the Fe-$d$ orbitals are split into $t_{2g}$ and $e_g$ groups, with the $t_{2g}$ group giving rise to the flat bands as shown in Figure 4b. Spin-orbit coupling (SOC) has a negligible effect on the bands near the Fermi energy) and neither does the choice of Hubbard $U$ as long as it is larger than a minimal value, in agreement with other similar Fe oxides(28).

The tight-binding (TB) model of the Fe-$t_{2g}$ states was constructed using Wannier90 functions (29) to investigate the origin of the flat bands. It was found that the Wannier Hamiltonian from the Fe-$t_{2g}$ states consist of contributions from both direct Fe-Fe electron hopping and indirect (assisted by Fe-O and Sb-O orbital overlap) electron hopping. Figure 4c shows a comparison between the actual $t_{2g}$ spectrum (which includes both direct and indirect electron hopping) and a spectrum that considers only direct hopping. It is found that the $d_{xz}$ and $d_{yz}$ states are slightly affected by indirect hopping and that the $d_{xy}$-$d_{xy}$ hopping probability (i.e. the sub-band energy width) is increased by 5 times. Hence, the direct Fe-Fe electron hopping probability, which is small, is enhanced by the indirect electron hopping coming from Fe-O and Sb-O states for the Fe $d_{xy}$ orbitals, and to a much lesser but significant degree for the $d_{xz}$ orbitals. Due to the only weakly broken 4-fold rotational symmetry parallel to $c$, the contribution of indirect electron hopping to the total electron hopping probability in the *xy* plane is largest. That is, the presence of a 4-fold rotational symmetry parallel to *c* implies zero out-of-plane electron hopping. In fact, it is much larger than the direct hopping, as is evidenced in Figure 4c. Furthermore, among the $d_{yz}$ and $d_{xz}$ orbitals, one of the sub-bands ($d_{yz}$) is significantly flatter than the other. The origin of this very flat sub-band can be attributed to the interference of electron hopping instead of its absence. Namely, the in-plane $d_{zx}$, $d_{yz}$ hoppings are close (in parameter space) to a Hamiltonian with a completely flat band in the $k_x - k_y$ plane. All the details for the theoretical calculations are provided in the Supporting Information.

**Physical Property Characterization**

*Magnetic Properties*

The zero-field-cooled (ZFC) and field-cooled (FC) temperature-dependent DC magnetization data were collected for each composition of the $Sr_{2-x}La_xFeSbO_6$ ($0 \leq x \leq 0.3$) double perovskite series under an applied field of $H = 0.1$ T and then plotted as magnetic susceptibility χ against temperature (Figure 5a). A divergence between the ZFC and FC curves is clearly observed for each composition after going through a local maximum in the magnetic susceptibility. This behavior marks the onset of magnetic ordering. Thus the DC magnetic susceptibility data, over the suitable temperature range (selected as the straight-line part of the 1/χ vs. T curves, marked in red in Figure 5b) for each composition were fitted to the Curie-Weiss law (χ = C/(T - θ) + χ₀), to yield the Curie constants *C*, Weiss temperatures θ and effective moments μ$_{eff}$ that are listed in Table 2. The field-dependent magnetization data were collected from each composition as well, at T = 300 K and 1.8 K, and are plotted as *M* against *H* ($-9 \leq H / T \leq 9$) in Figure 5c & d, respectively.

The undoped parent phase $Sr_2FeSbO_6$ ($x = 0$) has been reported to adopt long-range, A-type (i.e. type I) antiferromagnetic ordering based on low-temperature neutron powder diffraction

data.(12) The ZFC temperature-dependent magnetization data collected from our $Sr_2FeSbO_6$ shows an antiferromagnetic transition around $T_N \approx 33$ K, in good agreement with the literature. However, a novel, second transition ($T_A \approx 15$ K) is observed in every batch of this compound at lower temperature. Based on the structural refinement against the high-intensity high-resolution synchrotron XRD data, the presence of a minority phase in the parent compound is unlikely, and hence we believe that the transition at $T_A$ is intrinsic. This observation suggests that an additional feature apart from the well-established antiferromagnetic transition, for example, potentially some glassiness or frustration, is likely to present to a certain extent in the parent compound (Figure 5a & b). However, one of the diagnostic features of a typical spin-glass system (low-temperature field-dependent magnetization data displaced from the origin)(30) is absent in our *M* vs *H* plots (Figure 5c & d), which could be due to the nearly-fully-ordered character of our sample and the relatively low sensitivity of our *M* vs. *H* measurements. To further investigate these two transitions ($T_N$ and $T_A$), AC magnetic susceptibility data χ' was collected at various frequencies with a small DC field (10 Oe) applied. These data are plotted against temperature T in the inset in Figure 5a. The transitions are both visible: the one around $T_N \approx 33$ K is relatively resistant to the change in AC frequency, while the transition around $T_A \approx 15$ K is quite labile. This supports the conclusion that the transition at $T_N \approx 33$ K is attributed to long-range antiferromagnetic ordering, while the transition at $T_A \approx 15$ K should be attributed to a less well-defined transition in the system, possibly as a consequence of the small amount of anti-site disorder existing between B-cations in the parent compound. The coexistence of a magnetically ordered spin system and a spin-glass system for the parent compound has been postulated in a previous study.(12) Thus, in our case, if the lower temperature transition is assumed to be a spin-glass transition, then that is consistent with expectations that the glassiness vanishes upon La-doping as a result of complete ordering between B-site cations.

The La-doped materials $Sr_{2-x}La_xFeSbO_6$ ($x$ = 0.1, 0.2, 0.3) show a systematic increase in the magnitude of DC magnetic susceptibility, accompanied by a subtle but continuous shift of the local maximum to a lower temperature (Figure 5a & b). At 300 K, the isothermal magnetization as a function of applied field for the materials studied is linear and passes through the origin with positive slope, while the analogous data collected at 1.8 K exhibit hysteresis with the "loop" systematically opening up upon increasing electron doping (Figure 5c & d), all consistent with temperature-dependent magnetization data.

The Curie constant $C$ (Figure 6a), Weiss temperature $\theta$ (Figure 6b) and effective moment per formula unit, $\mu_{eff}$ (Figure 6c), extracted from the Curie-Weiss fits of DC magnetic susceptibility data over the same high temperature range (200 – 300 K), all progress linearly with respect to the La-doping level $x$. The Curie constant $C$ extracted from $Sr_2FeSbO_6$ ($x$ = 0) is 3.796(7) cm$^3$ K mol$^{-1}$, and hence yields an observed $\mu_{eff}$ value of 5.511 $\mu_B$/f.u., corresponding to ~93% of the expected spin-only value for an $S$ = 5/2 system (Figure 6a). Since $Fe^{3+}$ in this case undoubtedly exhibits a high-spin (h.s.) d$^5$ configuration, and upon formal reduction to $Fe^{2+}$ due to the $La^{3+}$–for–$Sr^{2+}$ substitution, the number of unpaired electrons should decrease independent of whether the $Fe^{2+}$ displays high-spin (h.s.), intermediate-spin (i.s.) or low-spin (l.s.) d$^6$ configuration (Figure 6d). If the overall observed effective moment $\mu_{eff}$ is a linear combination of $Fe^{3+}$ and $Fe^{2+}$ centers in our materials, an expected effective moment $\mu_{eff}$ can be calculated from the spin-only formula. The observed effective moment $\mu_{eff}$ (black solid line) is plotted together with the calculated ones (normalized to the value at 0 electron doping) for all three scenarios (red dotted line for h.s.+ l.s., green dotted line for h.s.+ i.s. and blue dotted line for h.s.+ h.s.) in Figure 6c. Our experimental data thus suggest a possibility of high-spin $d^5$ $Fe^{3+}$ plus intermediate-spin $d^6$ $Fe^{2+}$ scenario for our materials, even though $O^{2-}$ is a relatively weak field ligand, which in principle should favor the high-spin electronic configuration of a $3d$ transition metal like Fe. The intermediate-spin configuration of $Fe^{2+}$ ($S = 1$) is not commonly seen, but a few intermediate-spin iron(II) complexes have been reported in the literature(31–33). Another possible explanation for what may appear to be an unusual intermediate-spin $Fe^{2+}$ ion is, contrary to the case where all the spin values deduced are based on a localized spin picture, that in this flat-band electronic system, in which the doped electrons are in a complex band, the effective moment per doped electron is actually lower than expected from a combination of localized high-spin $d^5$ $Fe^{3+}$ plus localized high-spin $d^6$ $Fe^{2+}$. We also note that for all the materials studied here, the temperature dependence of the magnetic susceptibility deviates from strict Curie Weiss behavior on decreasing the temperature below about 150 K. We have no data that indicates why this is the case.

The Weiss temperature θ progresses from a large negative value to a small positive value with increasing La-dopant concentration (Figure 6b). This indicates that the antiferromagnetic coupling between ferromagnetically-ordered moments that dominates the bulk magnetic behavior of $Sr_2FeSbO_6$ (an A-type antiferromagnet) is significantly weakened by introducing the electrons. At the threshold composition of the single-phase solid solution studied here ($x$ = 0.3), the ferromagnetic coupling within the layers eventually outweighs the competing antiferromagnetic interactions between layers. Magnetic characterization at higher temperatures, which is not available in our laboratory, may be of interest in future studies designed to investigate the magnetism of these materials above room temperature.

*Heat Capacity*

Heat capacity data was collected from both end members of the $Sr_{2-x}La_xFeSbO_6$ ($0 \leq x \leq 0.3$) double perovskite series: the undoped parent phase ($x = 0$) at various external applied fields ($\mu_0H$ = 0, 3, 6, 9 T) and the threshold composition ($x = 0.3$) at zero applied field. Data were also collected for the non-magnetic analogue $Sr_2GaSbO_6$ at zero applied field over the temperature range $1.8 \leq T / K \leq 50$.

The total heat capacity $C_{total}$ data are plotted against temperature in Figure 7a. The peak observed in $C_{total}$ curves of the undoped parent phase $Sr_2FeSbO_6$ near $T \approx 33$ K is again seen to be relatively resistant to changes in the external applied field, corresponding to the phase transition associated with bulk long-range antiferromagnetic ordering observed in the temperature-dependent magnetization data. The lower temperature transition ($T \approx 15$ K) observed in the magnetic susceptibility cannot clearly be seen in the total heat capacity $C_{total}$ curve, consistent with the possibility that only a small amount of glassiness is present in the system due to the limited anti-site disorder. Furthermore, the feature around $T \approx 33$ K which is associated with the long-range antiferromagnetic ordering in the undoped parent phase $Sr_2FeSbO_6$ is absent in analogous total heat capacity data $C_{total}$ collected from the threshold composition $Sr_{1.7}La_{0.3}FeSbO_6$, $x = 0.3$. This is unusual and suggests that the phase transition due to long-range antiferromagnetic ordering has been significantly suppressed in the doped sample in spite of its significant drop in susceptibility near 30 K. The introduction of $La^{3+}$ dopant seems to severely interfere with the original long-range A-type antiferromagnetic ordering in the undoped material, hence yielding the disappearance of a signature λ-shaped transition in the thermodynamic data collected from the threshold composition. This peculiar feature postulates that the transition observed in the magnetic susceptibility of $Sr_{1.7}La_{0.3}FeSbO_6$ may not be long-range and its origin may be of future interest.

The total heat capacity of a magnetic material can be interpreted as the sum of electronic, phonon and magnetic contributions ($C_{total} = C_{electron} + C_{phonon} + C_{mag}$). Thus to isolate the magnetic contribution $C_{mag}$ to determine the magnetic entropy change $\Delta S_{mag}$, the heat capacity contributions due to electrons ($C_{electron}$) and phonons ($C_{phonon}$) need to be estimated and subtracted. To make this approximation, the heat capacity data collected from $Sr_2GaSbO_6$, a non-magnetic analogue of $Sr_2FeSbO_6$, has been rescaled here to minimize the non-magnetic differences between $Fe^{3+}$ and $Ga^{3+}$, hence serving as an approximation for $C_{phonon}$ of $Sr_2FeSbO_6$. $C_{electron}$ is effectively = 0 for insulating materials such as these (the resistivity of all the materials prepared in this study is too high to measure at ambient temperature and increases with decreasing temperature meaning that there are no conduction electrons present that would yield a significant $C_{electron}$). The resulting $C_p/T$ is plotted against temperature in Figure 7b. After subtracting the approximate phonon contributions, equivalent to rescaled $C_{total}/T$ of $Sr_2GaSbO_6$, an approximate magnetic contribution $C_{mag}/T$ of $Sr_2FeSbO_6$ can be extracted. Then the magnetic entropy change $\Delta S_{mag}$ can be estimated to yield a saturation value of ~ 7.14 J/mol/K which corresponds to ~ 48% of that expected for Heisenberg spin (for an $S = 5/2$ system ($Fe^{3+}$: high-spin $d^5$)) ordering, $R \ln (2S + 1) = 14.89$ J/mol/K. The experimental value is much closer to the Ising spin prediction $R \ln (2) = 5.76$ J/mol/K. We therefore tentatively conclude that the magnetic entropy released at the magnetic phase transition is associated with the bulk long-range ordering of spins in $Sr_2FeSbO_6$ that are closer to Ising-like than Heisenberg-like.

*Basic Optical Properties*

The diffuse reflectance spectra were collected from powder samples at ambient temperature in order to observe the evolution of the optical absorption behavior in the $Sr_{2-x}La_xFeSbO_6$ ($0 \leq x \leq 0.3$) double perovskite series. The pseudo-absorbance, transferred from the reflectance using the Kubelka-Munk function, is plotted against wavelength (nm) in Figure 8. The optical transitions were then analyzed based on Tauc plots via the direct-transition approach, as shown in the inset. The energy scales determined for these direct transitions, which for the La doped material do show a gradual decreasing trend (from 2.58 eV to 2.24 eV) with increasing La-dopant concentration. Thus the La-for-Sr substitution has a relatively small effect on the band gap, which is below the Fermi energy for the doped compositions, but the electrons in the narrow band do not yield a strong optical effect. Experimental work exploring more extreme conditions or different electron-dopants to overcome the current solubility threshold in this system may be of future interest.

**Conclusions**

Based on theoretical insights into the flat electronic bands expected to lie above the Fermi Energy in this double perovskite system, the parent phase $Sr_2FeSbO_6$ and its $La^{3+}$–for–$Sr^{2+}$ substituted phases were prepared via traditional high-temperature ceramic synthesis. Their structural and physical properties were characterized. $Sr_{2-x}La_xFeSbO_6$ ($0 \leq x \leq 0.3$) double perovskites crystallize in a monoclinic symmetry space group ($I2/m$) based on high-resolution synchrotron X-ray powder diffraction data collected at ambient temperature. A small amount of anti-site disorder exists in our $Sr_2FeSbO_6$ (~95% degree of B-cation ordering) and vanishes when the La-dopant is introduced into the system. These subtle structural changes are reflected in the physical properties of these materials. $Sr_2FeSbO_6$ possess two intrinsic transitions in magnetic susceptibility data. The first one at $T_N \approx 33$ K is attributed to the bulk long-range antiferromagnetic ordering. The origin of the second transition at $T_A \approx 15$ K is unclear but may be due to the small amount of spin-freezing/frustration in the system which disappears upon La-doping as a result of complete ordering between B-site cations in doped materials. The magnetization data collected from doped materials also suggest that the antiferromagnetic coupling used to dominate the bulk magnetic behavior is weakened by introducing $La^{3+}$ dopant into the system, and the dominant coupling between spins eventually becomes weakly ferromagnetic at the threshold composition $Sr_{1.7}La_{0.3}FeSbO_6$. The parameters extracted from the Curie-Weiss fit and the calculated band structures support the idea that in a localized moment picture some of the $d^5$ $Fe^{3+}$ centers are reduced to $d^6$ $Fe^{2+}$ centers upon $La^{3+}$–for–$Sr^{2+}$ substitution. Heat capacity data collected from two end members of our double perovskite series, $Sr_2FeSbO_6$ and $Sr_{1.7}La_{0.3}FeSbO_6$, completely agrees with collected magnetization data. In addition to the magnetic properties, the optical transitions experimentally evaluated were found to gradually decrease in energy scale with increasing La-dopant concentration, suggesting that the material does not fit into in a strictly "rigid band" scenario. The theoretical picture and experimental data thus suggest that the electron doped double perovskite $Sr_2FeSbO_6$ may be an unusual electronic system worthy of further study due to the potential electron occupancy of its flat electronic band. Furthermore, $Sr_2FeSbO_6$ is clearly not the sole example in this material family that with flat electronic bands near or at the fermi level, other candidates such as the Cr-analogue is of future interest as well.

## Materials and Methods

### Materials Synthesis

Approximately 1.0 g polycrystalline powder samples of the $Sr_{2-x}La_xFeSbO_6$ ($0 \leq x \leq 0.5$) double perovskites were prepared via traditional high-temperature ceramic synthesis. Stoichiometric amounts of $SrCO_3$ (Alfa Aesar, 99.99%), $La_2O_3$ (Alfa Aesar, 99.99%, dried at 900 °C), $Fe_3O_4$ (Alfa Aesar, 99.997%) and $Sb_2O_5$ (Alfa Aesar, 99.998%) were thoroughly ground together by using an agate mortar and pestle, and then transferred into an alumina crucible for calcinations in air. These reaction mixtures were first slowly (1 °C/min) ramped up to 1000 °C and held overnight to decompose the carbonate, and then were annealed at 1500 °C (temperature raised at 3 °C/min) for 2 periods of 48 hours with intermittent grindings. A powder sample of $Sr_2GaSbO_6$, as the non-magnetic analogue of undoped parent phase $Sr_2FeSbO_6$, was prepared as previously described(34) to provide a non-magnetic phonon standard for the heat capacity measurements.

**X-ray Powder Diffraction Measurements and Refinements**

The reaction progress was monitored using laboratory X-ray powder diffraction data collected at room temperature on a Bruker D8 FOCUS diffractometer (Cu Kα) over a 2θ range between 5° and 70°. Once the reactions were deemed finished, laboratory XRD data with much better statistical significance, covering a 2θ range between 5° and 110°, were collected from each sample to establish the sample purity. The high-resolution synchrotron X-ray powder diffraction data were acquired at Beamline 11-BM at the Advanced Photon Source at Argonne National Laboratory at a wavelength of ~ 0.4590 Å. Detailed structural refinements were performed by the Rietveld method(35) using the GSAS-II program, with crystal structures visualized by the VESTA program.

**Magnetization and Thermodynamic Measurements**

The magnetization data were collected using the ACMS II and VSM options of a Quantum Design Physical Property Measurement System (PPMS). Temperature-dependent magnetization ($M$) data were collected under an applied external field ($H$) of 1000 Oe. Magnetic susceptibility is defined as M/H. Field-dependent isothermal magnetization data between $H$ = 90000 Oe and – 90000 Oe were collected at $T$ = 300 K and 1.8 K. Heat capacity was measured using a standard relaxation method in the PPMS over the temperature range 1.8 to 50 K.

**Optics Measurements**

The diffuse reflectance spectra were collected from a powder sample of each composition at ambient temperature on a Cary 5000i UV-VIS-NIR spectrometer equipped with an internal DRA-2500 integrating sphere. The data were converted from reflectance to pseudo absorbance via applying the Kubelka–Munk function, and band gaps were calculated from Tauc plots by using a direct transition equation.(36)

**Theoretical Calculations**

Density-functional-theory-based (DFT) first principles calculations were performed using the projector augmented-wave (PAW) method as implemented in the VASP code.(37, 38) The generalized gradient approximation (GGA) was employed to deal with electron correlation.(39) Reciprocal space integrations were completed over a 7×7×5 Monkhorst-Pack *k*-point mesh. The energy cutoff was chosen to be 1.5 times as large as the values recommended for the relevant PAW pseudopotentials. Spin-orbit coupling (SOC) had an insignificant effect on the bands, and we therefore ignored it in our analysis. The Hubbard U was set to 3eV, as in other similar iron-oxides.(28) To study the flat-band properties of the compound, we obtained a real space tight-binding Hamiltonian using the maximally localized Wannier functions (MLWFs) approach.(29)

**Supporting Information**

The Supporting Information is available from the publisher or from the authors upon valid request.

**Acknowledgements**

The experimental research was primarily done at Princeton University, supported by the US Department of Energy Division of Basic Energy Sciences, through the Institute for Quantum Matter, Grant No. DE-SC0019331. The theory work was supported by the European Research Council (ERC) under the European Union's Horizon 2020 research and innovation programme (grant agreement No. 101020833), the ONR Grant No. N00014-20-1-2303, Simons Investigator

Grant No. 404513, the Gordon and Betty Moore Foundation through the EPiQS Initiative, Grant GBMF11070 and Grant No. GBMF8685, NSF-MRSEC Grant No. DMR-2011750, BSF Israel US foundation Grant No. 2018226, and the Princeton Global Network Funds. Use of the Advanced Photon Source at Argonne National Laboratory was supported by the U. S. Department of Energy, Office of Science, Office of Basic Energy Sciences, under Contract No. DE-AC02-06CH11357.

**References**

1. P. Schiffer, A. P. Ramirez, W. Bao, S.-W. Cheong, Low Temperature Magnetoresistance and the Magnetic Phase Diagram of $La_{1-x}Ca_xMnO_3$. *Phys Rev Lett* **75**, 3336–3339 (1995).

2. J. G. Bednorz, K. A. Müller, Possible highT c superconductivity in the Ba−La−Cu−O system. *Zeitschrift für Physik B Condensed Matter* **64**, 189–193 (1986).

3. R. E. Cohen, Origin of ferroelectricity in perovskite oxides. *Nature* **358**, 136–138 (1992).

4. P. Lacorre, F. Goutenoire, O. Bohnke, R. Retoux, Y. Lallgant, Designing fast oxide-ion conductors based on $La_2Mo_2O_9$. *Nature* **404**, 856–858 (2000).

5. R. Asahi, T. Morikawa, T. Ohwaki, K. Aoki, Y. Taga, Visible-Light Photocatalysis in Nitrogen-Doped Titanium Oxides. *Science* **293**, 269–271 (2001).

6. S. Vasala, M. Karppinen, $A_2B′B″O_6$ perovskites: A review. *Progress in Solid State Chemistry* **43**, 1–36 (2015).

7. M. T. Anderson, K. B. Greenwood, G. A. Taylor, K. R. Poeppelmeier, B-cation arrangements in double perovskites. *Progress in Solid State Chemistry* **22**, 197–233 (1993).

8. G. King, P. M. Woodward, Cation ordering in perovskites. *J Mater Chem* **20**, 5785–5796 (2010).

9. G. Blasse, New compounds with perovskite-like structures. *Journal of Inorganic and Nuclear Chemistry* **27**, 993–1003 (1965).

10. P. D. Battle, T. C. Gibb, A. J. Herod, J. P. Hodges, Sol-gel synthesis of the magnetically frustrated oxides $Sr_2FeSbO_6$ and $SrLaFeSnO_6$. *J Mater Chem* **5**, 75–78 (1995).

11. S.-O. Lee, T. Y. Cho, S.-H. Byeon, Magnetic Property of the Perovskite Fe(III) Oxide. *Bull Korean Chem Soc* **18**, 91–97 (1997).

12. E. J. Cussen, J. F. Vente, P. D. Battle, T. C. Gibbb, Neutron diffraction study of the influence of structural disorder on the magnetic properties of $Sr_2FeMO_6$ (M=Ta, Sb). *J Mater Chem* **7**, 459–463 (1997).

13. N. Kashima, K. Inoue, T. Wada, Y. Yamaguchi, Low temperature neutron diffraction studies of $Sr_2FeMO_6$ (M=Nb, Sb). *Applied Physics A* **74**, 805–807 (2002).

14. A. Faik, J. M. Igartua, E. Iturbe-Zabalo, G. J. Cuello, A study of the crystal structures and the phase transitions of $Sr_2FeSbO_6$, $SrCaFeSbO_6$ and $Ca_2FeSbO_6$ double perovskite oxides. *J Mol Struct* **963**, 145–152 (2010).

15. L. Jin, *et al.*, Magnetic cations doped into a Double Perovskite Semiconductor. *J Mater Chem C Mater* **10**, 3232–3240 (2022).

16. N. Regnault, *et al.*, Catalogue of flat-band stoichiometric materials. *Nature* **603**, 824–828 (2022).

17. F. Xie, Z. Song, B. Lian, B. A. Bernevig, Topology-Bounded Superfluid Weight in Twisted Bilayer Graphene. *Phys Rev Lett* **124**, 167002 (2020).

18. V. Peri, Z. Song, B. A. Bernevig, S. D. Huber, Fragile Topology and Flat-Band Superconductivity in the Strong-Coupling Regime. *Phys Rev Lett* **126**, 027002 (2021).

19. C. Z. Chang, *et al.*, Experimental observation of the quantum anomalous Hall effect in a magnetic topological Insulator. *Science* **340**, 167–170 (2013).

20. C. Z. Chang, *et al.*, High-precision realization of robust quantum anomalous Hall state in a hard ferromagnetic topological insulator. *Nat Mater* **14**, 473–477 (2015).

21. M. Serlin, *et al.*, Intrinsic quantized anomalous Hall effect in a moiré heterostructure. *Science* **367**, 900–903 (2020).

22. D. Călugăru, *et al.*, General construction and topological classification of crystalline flat bands. *Nat Phys* **18**, 185–189 (2022).

23. Y. Cao, *et al.*, Correlated insulator behaviour at half-filling in magic-angle graphene superlattices. *Nature* **556**, 80–84 (2018).

24. R. Bistritzer, A. H. MacDonald, Moiré bands in twisted double-layer graphene. *Proc. Natl Acad. Sci. USA* **108**, 12233–12237 (2011).

25. Y. Cao, *et al.*, Unconventional superconductivity in magic-angle graphene superlattices. *Nature* **556**, 43–50 (2018).

26. R. D. Shannon, Revised effective ionic radii and systematic studies of interatomic distances in halides and chalcogenides. *Acta Crystallographica Section A* **32**, 751–767 (1976).

27. L. Jin, *et al.*, Structure and properties of the $Sr_2In_{1-x}Sn_xSbO_6$ double perovskite. *J Solid State Chem* **314**, 123355 (2022).

28. H. Hsu, K. Umemoto, M. Cococcioni, R. M. Wentzcovitch, The Hubbard U correction for iron-bearing minerals: A discussion based on $(Mg,Fe)SiO_3$ perovskite. *Physics of the Earth and Planetary Interiors* **185**, 13–19 (2011).

29. A. A. Mostofi, *et al.*, An updated version of wannier90: A tool for obtaining maximally-localised Wannier functions. *Comput Phys Commun* **185**, 2309–2310 (2014).

30. J. E. Page, M. A. Hayward, CaMn$_{0.5}$Ir$_{0.5}$O$_{2.5}$: An Anion-Deficient Perovskite Oxide Containing Ir$^{3+}$. *Inorg Chem* **58**, 8835–8840 (2019).

31. M.-S. Liao, J. D. Watts, M.-J. Huang, Electronic Structure of Some Substituted Iron(II) Porphyrins. Are They Intermediate or High Spin? *J Phys Chem A* **111**, 5927–5935 (2007).

32. J. F. Kirner, W. Dow, W. Robert. Scheidt, Molecular stereochemistry of two intermediate-spin complexes. Iron(II) phthalocyanine and manganese(II) phthalocyanine. *Inorg Chem* **15**, 1685–1690 (2002).

33. J. C. Ott, H. Wadepohl, M. Enders, L. H. Gade, Taking Solution Proton NMR to Its Extreme: Prediction and Detection of a Hydride Resonance in an Intermediate-Spin Iron Complex. *J Am Chem Soc* **140**, 17413–17417 (2018).

34. L. Jin, *et al.*, Ferromagnetic Double Perovskite Semiconductors with Tunable Properties. *Advanced Science* **9**, 2104319 (2022).

35. H. Rietveld, A profile refinement method for nuclear and magnetic structures. *J Appl Crystallogr* **2**, 65–71 (1969).

36. A. Y. Tarasova, *et al.*, Electronic structure and fundamental absorption edges of KPb$_2$Br$_5$, K$_{0.5}$Rb$_{0.5}$Pb$_2$Br$_5$, and RbPb$_2$Br$_5$ single crystals. *Journal of Physics and Chemistry of Solids* **73**, 674–682 (2012).

37. G. Kresse, J. Furthmüller, Efficient iterative schemes for ab initio total-energy calculations using a plane-wave basis set. *Phys Rev B* **54**, 11169–11186 (1996).

38. G. Kresse, D. Joubert, From ultrasoft pseudopotentials to the projector augmented-wave method. *Phys Rev B* **59**, 1758–1775 (1999).

39. J. P. Perdew, Y. Wang, Accurate and simple analytic representation of the electron-gas correlation energy. *Phys Rev B* **45**, 13244–13249 (1992).

**Table 1.** Structural parameters and crystallographic positions from the refinement of synchrotron X-ray powder diffraction data collected from $Sr_{1.9}La_{0.1}FeSbO_6$ at 300 K.

| Atoms | x/a | y/b | z/c | S.O.F. | $U_{iso}$ equiv. (Å$^2$) |
|---|---|---|---|---|---|
| Sr1 | 0.49917(16) | 0 | 0.25004(5) | 0.95 | 0.00416 |
| La1 | 0.49917(16) | 0 | 0.25004(5) | 0.05 | 0.00416 |
| Fe1 | 0 | 0 | 0 | 1 | 0.0009 |
| Sb1 | 0 | 0 | 0.5 | 1 | 0.0004 |
| O1 | -0.0438(5) | 0 | 0.25219(31) | 1 | 0.0039 |
| O2 | 0.25276(31) | 0.25307(31) | 0.0230(3) | 1 | 0.0097 |

$Sr_{1.9}La_{0.1}FeSbO_6$ space group $I2/m$ (#12)
Formula weight: 453.969 g mol$^{-1}$, $Z = 2$
$a = 5.62104(1)$ Å, $b = 5.59656(1)$ Å, $c = 7.90033(1)$ Å, $\beta = 89.9969(6)$ °, $V = 248.5326(8)$ Å$^3$
Radiation source: Beamline 11-BM at a wavelength of ~ 0.4590 Å
Temperature: 300 K
$wR = 13.791$ %; $GOF = 1.95$

Note: The S.O.F. of the A-site cations (Sr and La) are assigned based on the original stoichiometry ratio of elements put in during the sample preparation.

**Table 2.** The Curie constant and Weiss temperature extracted from the fitting of paramagnetic susceptibility to the Curie-Weiss law, the observed effective moment per formula unit, and the calculated effective moment per formula unit predicted based on the spin-only formula for each composition of the $Sr_{2-x}La_xFeSbO_6$ series.

| Composition | Curie constant $C$ (cm$^3$ K mol$^{-1}$) | Observed effective moment per formula unit $\mu_{eff.obs}$ ($\mu_B$/f.u.) | Calculated effective moment per formula unit $\mu_{eff.cal}$ ($\mu_B$/f.u.) (h.s. d$^5$ Fe$^{3+}$ and l.s. d$^6$ Fe$^{2+}$) | Calculated effective moment per formula unit $\mu_{eff.cal}$ ($\mu_B$/f.u.) (h.s. d$^5$ Fe$^{3+}$ and i.s. d$^6$ Fe$^{2+}$) | Calculated effective moment per formula unit $\mu_{eff.cal}$ ($\mu_B$/f.u.) (h.s. d$^5$ Fe$^{3+}$ and h.s. d$^6$ Fe$^{2+}$) | Weiss temperature $\theta$ (K) |
|---|---|---|---|---|---|---|
| $Sr_{2-x}La_xFeSbO_6$ | | | | | | |
| $x = 0$ | 3.796(7) | 5.511 | 5.916 | 5.916 | 5.916 | −169.2(6) |
| $x = 0.1$ | 3.510(7) | 5.299 | 5.324 | 5.607 | 5.814 | −126.0(6) |
| $x = 0.2$ | 3.019(6) | 4.914 | 4.733 | 5.299 | 5.713 | −54.6(5) |
| $x = 0.3$ | 2.839(5) | 4.766 | 4.141 | 4.990 | 5.611 | +5.0(5) |

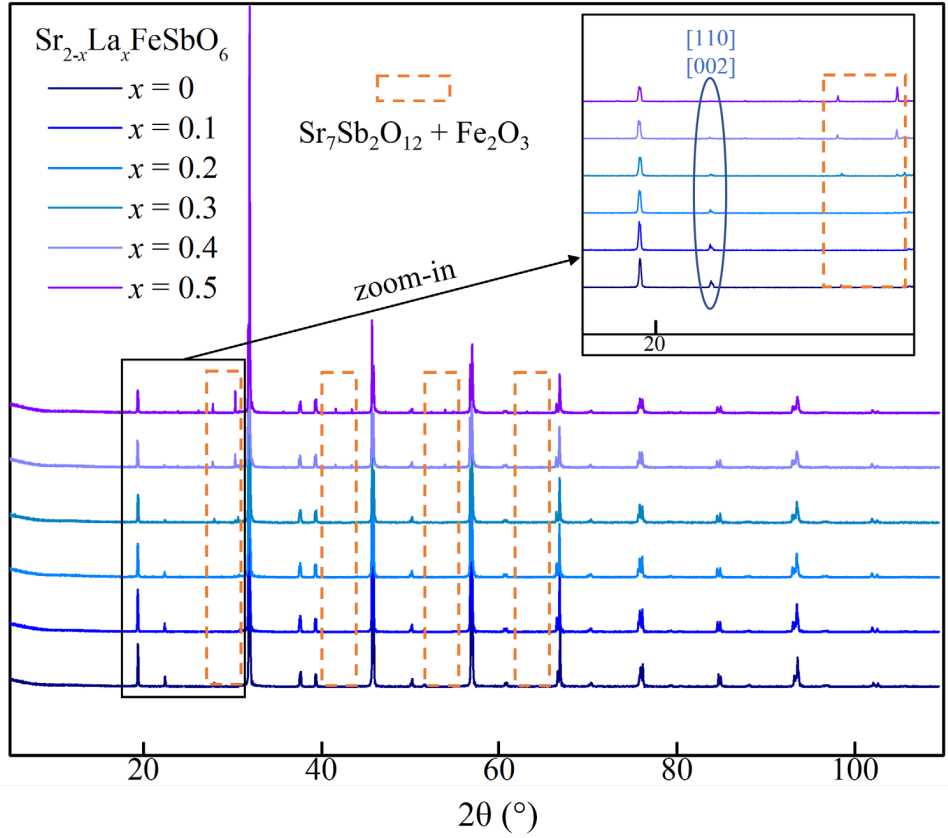

**Figure 1.** Stacked lab X-ray powder diffraction patterns for compositions of the $Sr_{2-x}La_xFeSbO_6$ series, with orange boxes highlighting the growth of peaks representing the impurity phases $Sr_7Sb_2O_{12}$ and $Fe_2O_3$. Inset: low angle area zoomed in.

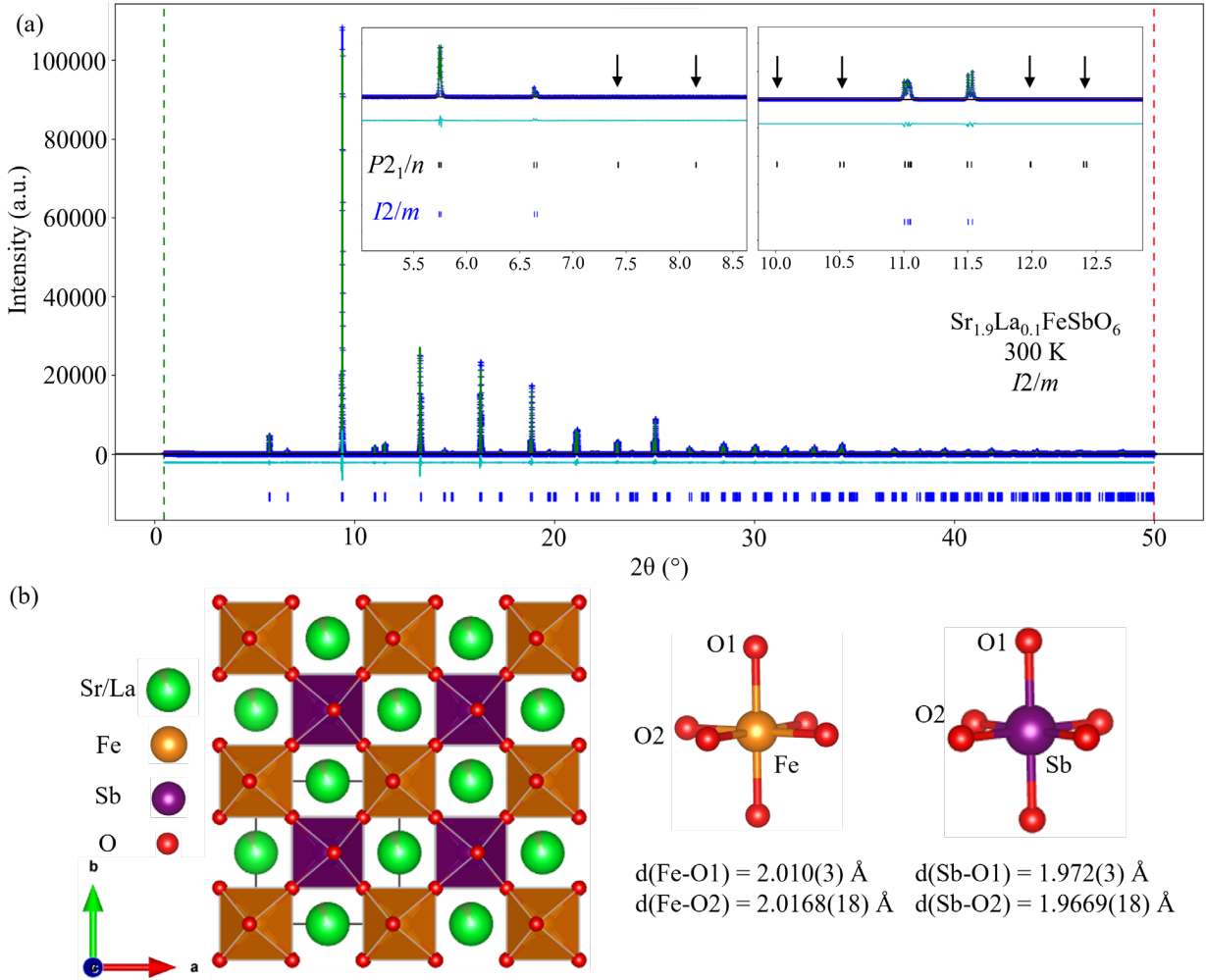

**Figure 2.** (a) Observed (blue), calculated (green), and difference (cyan) plots from the Rietveld refinement of $Sr_{1.9}La_{0.1}FeSbO_6$ (space group $I2/m$) against the synchrotron X-ray powder diffraction (SXRD) data collected at ambient temperature; Insets show in detail the absence of primitive Bragg reflections associated with the $P2_1/n$ symmetry in our data (positions marked by arrows). (b) The structural model for $Sr_{1.9}La_{0.1}FeSbO_6$ together with selected bond lengths for the $FeO_6$ octahedron and the $SbO_6$ octahedron.

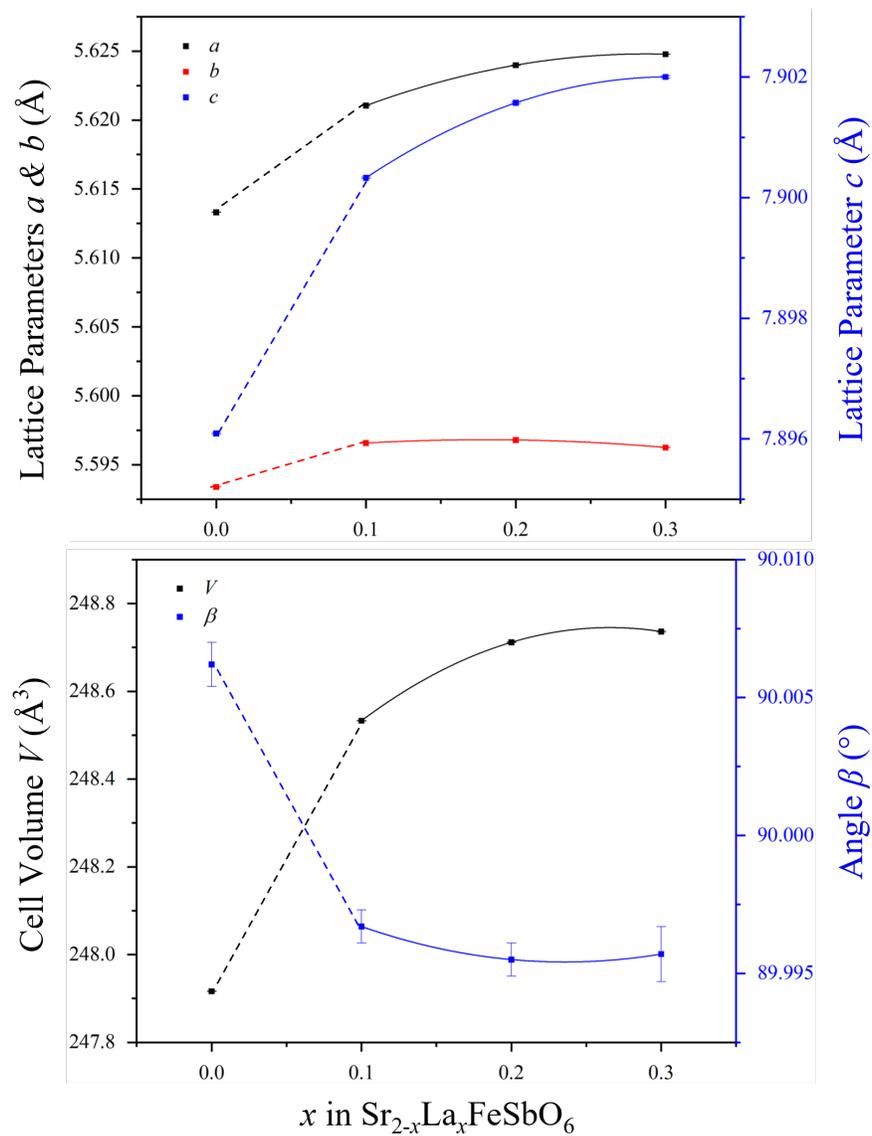

**Figure 3.** Refined lattice parameter *a*, *b*, *c*, cell volume *V* and angle *β* plotted for each composition of the $Sr_{2-x}La_xFeSbO_6$ series.

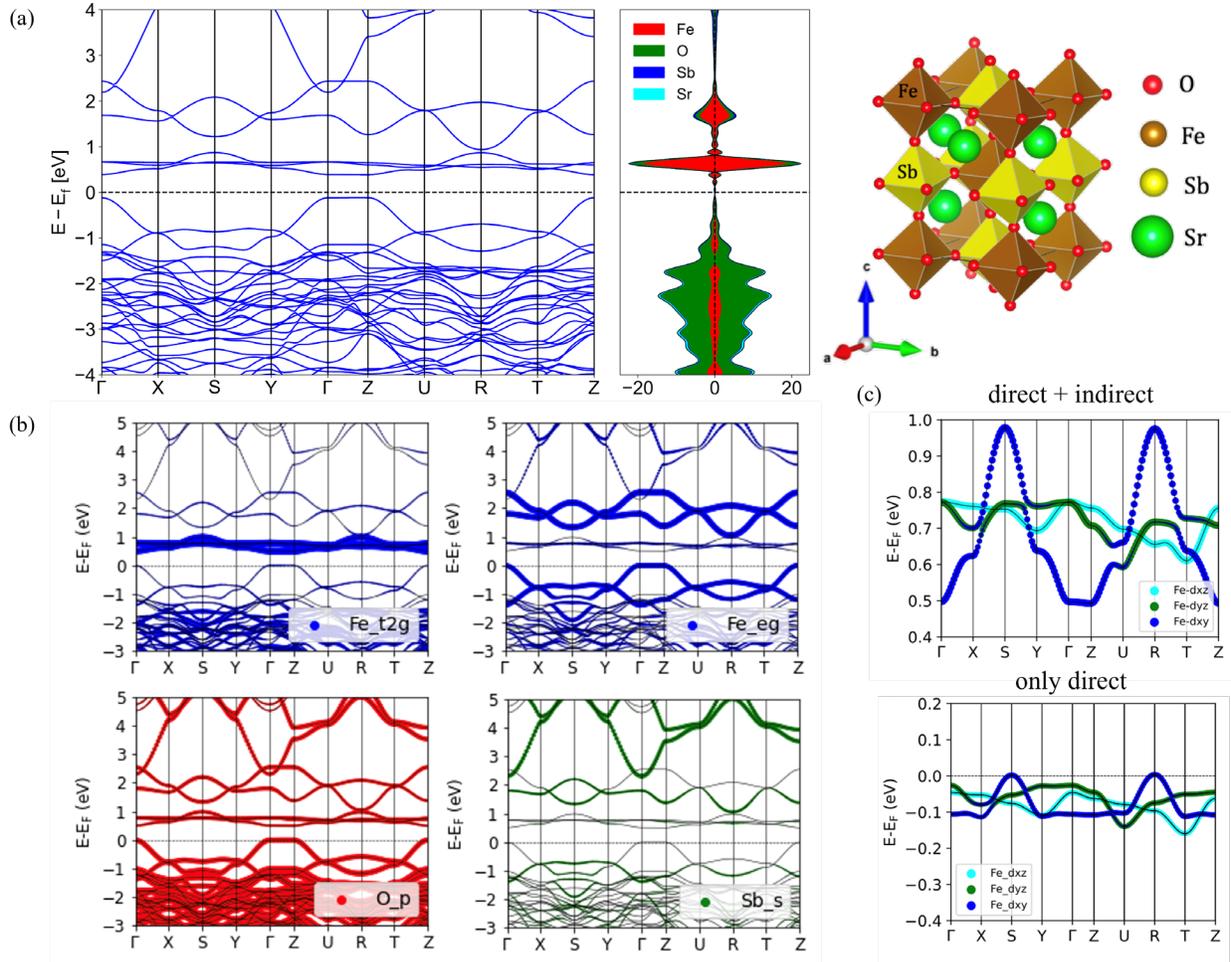

**Figure 4.** (a) The calculated band structure and electronic density of states (DOS) for double perovskite $Sr_2FeSbO_6$; (b) orbital character of the states close to the Fermi level; (c) band structure of the Fe-$t_{2g}$ states considering direct + indirect electron hopping and only direct electron hopping.

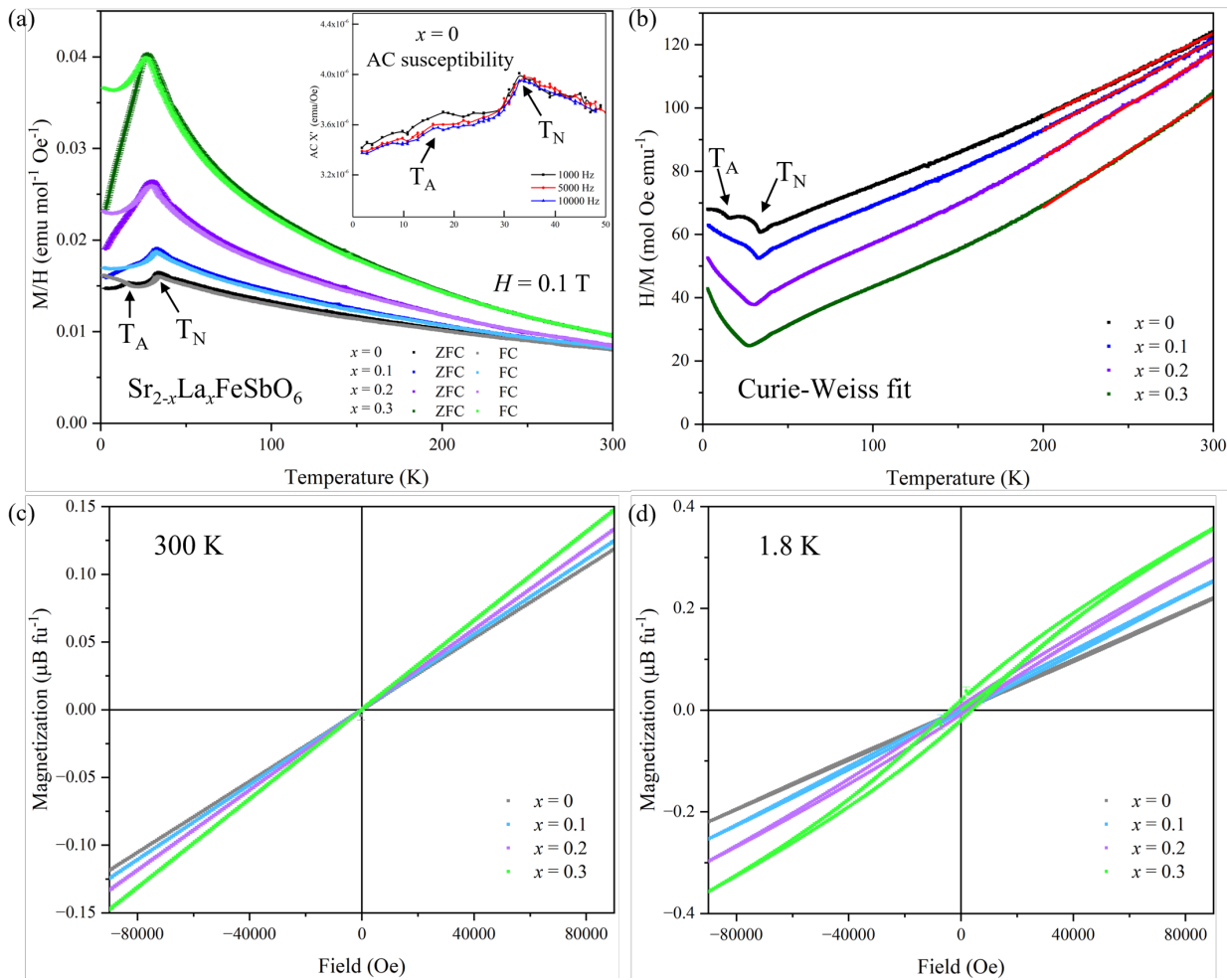

**Figure 5.** (a) The zero-field-cooled (ZFC) and field-cooled (FC) temperature-dependent DC magnetization data collected under an applied field of $H = 0.1$ T; Inset shows the AC magnetic susceptibility data $\chi'$ collected from $Sr_2FeSbO_6$ at various frequencies with a small DC field (10 Oe) applied; (b) The inverse of magnetic susceptibility $1/\chi$ plotted against temperature T, with the straight-line part (marked in red) fitted to the Curie-Weiss law; The field-dependent magnetization data collected at (c) 300 K and (d) 1.8 K for each composition of the $Sr_{2-x}La_xFeSbO_6$ ($0 \leq x \leq 0.3$) double perovskite series.

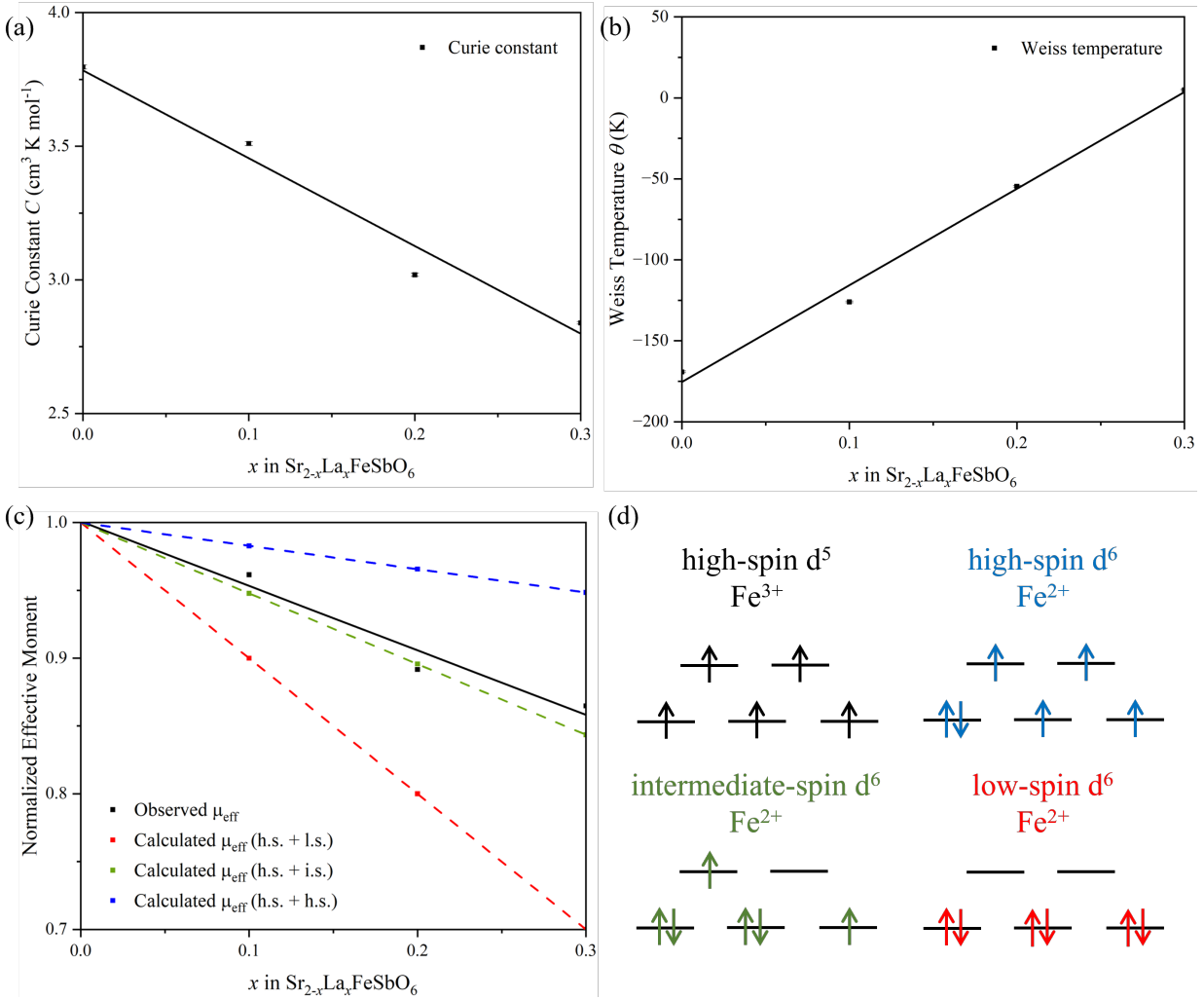

**Figure 6.** (a) The Curie constant and (b) the Weiss temperature extracted from the fitting of paramagnetic susceptibility to the Curie-Weiss law; (c) The normalized observed effective moment per formula unit plotted against the normalized effective moment per formula unit calculated from the spin-only contribution based on three possible combinations of spin configurations for each composition of the $Sr_{2-x}La_xFeSbO_6$ ($0 \leq x \leq 0.3$) double perovskite series; (d) Possible spin configurations of $Fe^{3+}$ and $Fe^{2+}$ centers.

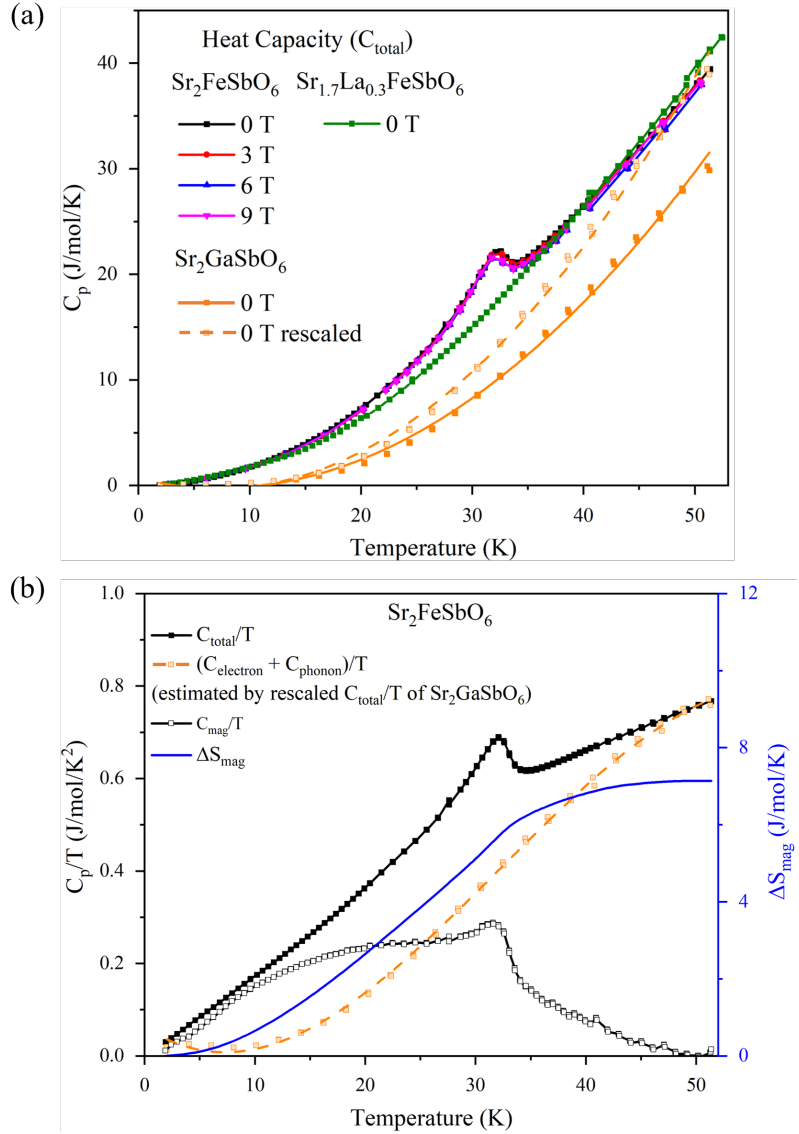

**Figure 7.** (a) The total heat capacity $C_{total}$ plotted against temperature for $Sr_2FeSbO_6$ at $H$ = 0, 3, 6, 9 T field applied, $Sr_{1.7}La_{0.3}FeSbO_6$ and $Sr_2GaSbO_6$ at zero field applied; (b) All relevant $C_p/T$ plotted against temperature T to extract the magnetic contribution to heat capacity and entropy change in $Sr_2FeSbO_6$.

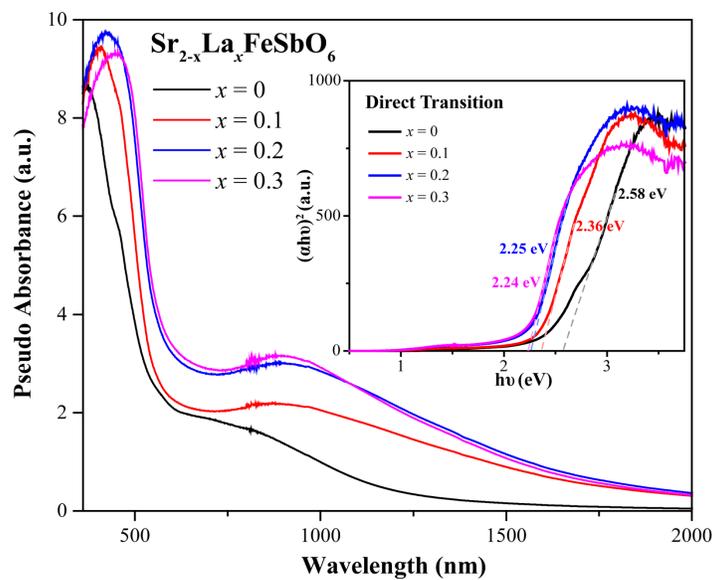

**Figure 8.** The diffuse reflectance spectra collected from the $Sr_{2-x}La_xFeSbO_6$ ($0 \leq x \leq 0.3$) double perovskite series with the energy scales of optical transitions extracted from Tauc plots obtained by using a direct transition equation.